\documentclass[twocolumn,showpacs,aps,prl,superscriptaddress,unsortedaddress]{revtex4}
\usepackage{graphicx}
\usepackage{epsf}
\newcommand{\etal}{{\it et al.}}

\begin{document}

\title{Protected nodes and the collapse of the Fermi arcs  in high T$_c$ cuprates}

\author{A. Kanigel}
\affiliation{Department of Physics, University of Illinois at Chicago, Chicago, IL 60607}
\author{U. Chatterjee}
\affiliation{Department of Physics, University of Illinois at Chicago, Chicago, IL 60607}
\author{M. Randeria}
\affiliation{Department of Physics, The Ohio State University, Columbus, OH  43210}
\author{M. R. Norman}
\affiliation{Materials Science Division, Argonne National Laboratory, Argonne, IL 60439}
\author{S. Souma}
\affiliation{Department of Physics, University of Illinois at Chicago, Chicago, IL 60607}
\author{M. Shi}
\affiliation{Department of Physics, University of Illinois at Chicago, Chicago, IL 60607}
\affiliation{Swiss Light Source, PSI, CH-5232 Villigen, Switzerland}
\author{Z. Z. Li}
\author{H. Raffy}
\affiliation{Laboratorie de Physique des Solides, Universite Paris-Sud, 91405 Orsay Cedex, France}
\author{J. C. Campuzano}
\affiliation{Department of Physics, University of Illinois at Chicago, Chicago, IL 60607}
\affiliation{Materials Science Division, Argonne National Laboratory, Argonne, IL 60439}

\begin{abstract}
Angle resolved photoemission on underdoped
Bi$_2$Sr$_2$CaCu$_2$O$_8$ reveals that the magnitude and d-wave anisotropy of the 
superconducting state energy gap are independent of temperature all the way up to T$_c$.
This lack of $T$ variation of the entire ${\bf k}$-dependent gap is in marked contrast to
mean field theory.   
At T$_c$ the point nodes of the d-wave gap abruptly expand into finite length ``Fermi arcs''.  
This change occurs within the width of the resistive transition, and thus the Fermi arcs are not simply thermally 
broadened nodes but rather a unique signature of the pseudogap phase.  
\end{abstract}

\date{\today }
\maketitle

We present in this Letter angle resolved photoemission spectroscopy (ARPES) data on the energy gap in the superconducting (SC) and
pseudogap phases of the underdoped high $T_c$ superconductor Bi$_2$Sr$_2$CaCu$_2$O$_8$ (Bi2212). 
Reduced $T_c$ samples which lie in between the optimally doped, highest $T_c$ material 
and the undoped Mott insulator are called ``underdoped''. Such samples exhibit an unusual normal-state pseudogap,
whose signature in ARPES \cite{Our_review,Shen_review} is a loss of spectral weight in parts of ${\bf k}$-space, leading to low energy electronic excitations 
which live on disconnected ``Fermi arcs'' \cite{Norman_Nature98}. Both the SC gap below $T_c$ and the pseudogap above $T_c$ 
are anisotropic gaps in the single-particle excitation spectrum, but their relationship is not well understood. 
Our goal is to gain insight into how the SC gap, with its d-wave 
anisotropy \cite{Our_review,Shen_review} at low $T$, changes as a function of temperature and evolves into 
the anisotropic pseudogap upon heating through $T_c$.    

Our first result is that the magnitude and anisotropy of the d-wave SC energy gap  
is essentially $T$-independent for all $T \le T_c$. This behavior is completely
different from a mean field description of a d-wave superconductor. Within mean field theory the anisotropy would be $T$-independent, but the 
gap magnitude, proportional to the order parameter, would be suppressed with increasing $T$ and vanish at $T_c$.
Remarkably, our data show that the ${\bf k}$-dependent SC gap in underdoped cuprates does not even know about the scale of $T_c$.

Second, there are four point nodes at which the energy gap vanishes for all $T < T_c$, but above $T_c$
each point node abruptly expands into a gapless ``Fermi arc'' of finite extent. We show
that this remarkable change occurs within the width of the resistive transition at $T_c$. 
The abrupt change from point nodes to gapless arcs is not just thermal smearing, 
but rather it is closely connected with the loss of superconducting order.
 
We present ARPES data on two underdoped Bi2212 films, one near optimality ($T_c = 80$K) and the other more underdoped ($T_c = 67$K),
both of which exhibit a significant pseudogap. The films, prepared by RF sputtering on a SrTiO$_3$ substrate \cite{Raffy_T*}, 
exhibit only very weak superstructure replicas which simplifies the data analysis. The measurements were carried out at the Synchrotron Radiation Center, Wisconsin, using a Scienta R4000 analyzer with 22eV photons, and momentum cuts and polarization parallel to $(0,0) \rightarrow (\pi,0)$.  The energy resolution was 20 meV (FWHM) with a ${\bf k}$-resolution 
along the multiplex direction of 0.0055 $\AA^{-1}$. The measured ARPES intensity is proportional to the spectral function times the Fermi function. 
To determine gap values, we remove the effects of the Fermi function from the data following two procedures: 
(i) symmetrization (adding spectra at $\pm$E) \cite{Norman_Nature98} and (ii) division of the spectra by a resolution-broadened Fermi function (obtained from polycrystalline Au in contact with the sample). Both methods gave identical gap estimates within the error bars quoted below.  

\begin{figure*}
\begin{center}
\includegraphics[width=13cm]{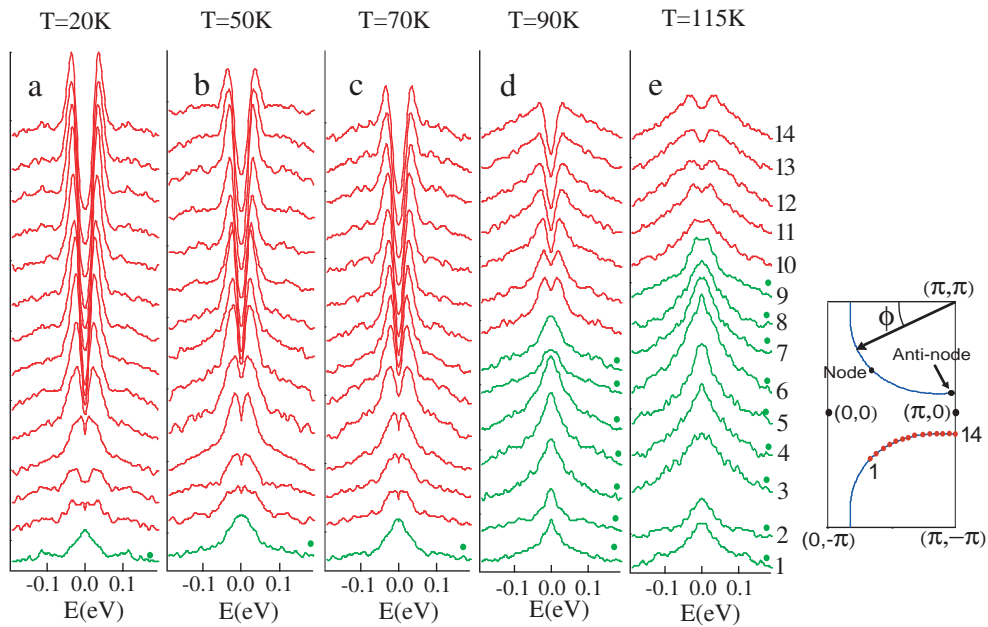}
\end{center}
\caption{Symmetrized EDCs for the T$_c$=80K sample at various temperatures. The different EDCs cover the whole Fermi surface from the node (lowest curve) up to the antinode (uppermost curve). Gapless spectra are indicated by solid dots.  The Fermi surface and location of the
various EDCs are shown in the far right panel.}
\label{raw_data}
\end{figure*}

In Fig.~\ref{raw_data} we show symmetrized spectra for the $T_c = 80$K sample at various temperatures for 14 angles along the Fermi surface \cite{KF}, indicated in the Brillouin zone (far right panel). The spectra in panels {\textbf a} (20K) , {\textbf b} (50K) and {\textbf c} (70K) are in the SC state. A gapless (symmetrized) spectrum has a maximum at zero energy, while gapped spectra show a minimum at zero 
with coherent peaks at ($\pm$) the gap value \cite{GAP}. One can clearly see the variation of the SC gap along the Fermi surface. It decreases monotonically from its maximum value at the antinode (top curve, point 14 on the zone edge) eventually vanishing at the node (bottom curve, point 1 on the zone diagonal). 
    
Our main conclusions for the underdoped SC state are:
(1)  There are sharp quasiparticle peaks for all $T < T_c$
which indicate coherent, long-lived, electronic excitations in the SC state.
(2) Both the magnitude of the SC gap and its ${\bf k}$-variation are 
independent of $T$ in the SC state. (3) In particular, the point node in the SC gap 
is protected for all $T < T_c$. We emphasize that these features of the SC gap
are observed all the way to $70K$ in a $T_c = 80$K sample. Even at 
$T = 0.87 T_c$, there is no indication in the ${\bf k}$-dependent SC gap, that the system is about to lose its superconductivity at $T_c$.
 
\begin{figure}
\begin{center}
\includegraphics[width=7cm]{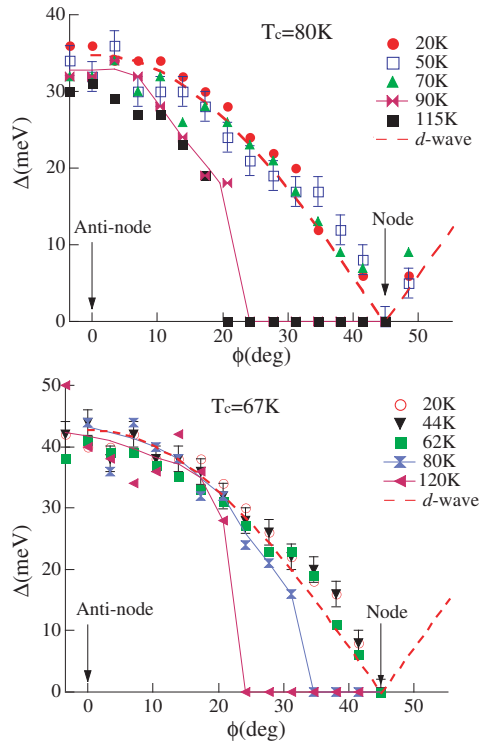}
\end{center}
\caption{The gap $\Delta$ as function of $\phi$, the Fermi surface angle measured from the zone boundary. We show data for two samples at five different temperatures. The dashed line is the expected angular dependence of a d-wave gap.}
\label{Delta_vs_Phi}
\end{figure}

To elucidate the $T$ (in)dependence of various features in detail, we present a quantitative analysis of the data. In Fig.~\ref{Delta_vs_Phi} we plot the gap $\Delta$ as function of the angle $\phi$ along the Fermi surface for the two samples.  The gap is defined as the separation of the peak from zero energy (Fig.~\ref{raw_data}).
For the three temperatures $T < T_c$, the SC gap is  \emph{constant in magnitude}, being maximum at the antinode at $\phi = 0^\circ$ and 
decreasing to zero at the node at $\phi = 45^\circ$, consistent with the d-wave form $\vert \cos k_x - \cos k_y\vert$ within error bars.

\begin{figure}
\begin{center}
\includegraphics[width=7cm]{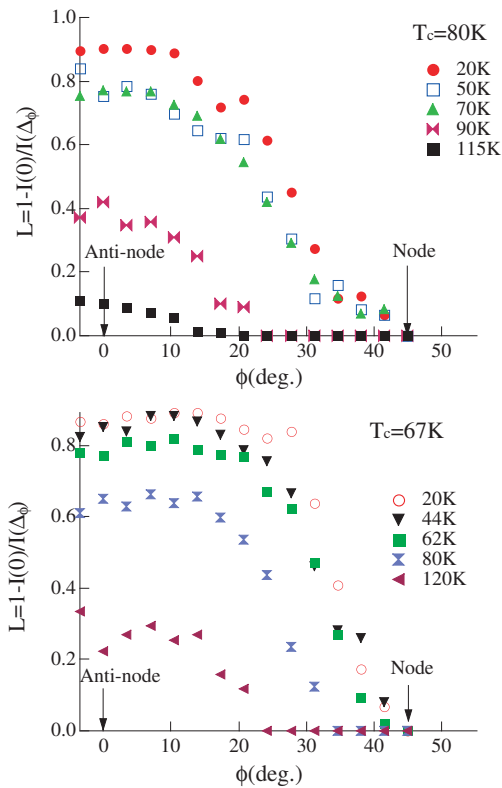}
\end{center}
\caption{Loss of zero energy spectral weight $L(\phi)=1-I(0)/I(\Delta_{\phi})$ as a function of $\phi$, the Fermi surface angle measured from the zone boundary (see text). The two panels show results at five different temperatures for two samples. }
\label{L_vs_Phi}
\end{figure}

As evident from Fig.~\ref{raw_data}, there are non-trivial lineshape changes \cite{Norman_PRB98}
with temperature. Thus, in addition to characterizing the data in terms of the energy gap $\Delta$, 
we also look at the $T$-dependent changes in the zero energy intensity relative to that at $\Delta$. 
We use a model-independent way of characterizing \cite{Nature_Physics} the loss of low-energy spectral weight defined by $L(\phi)=1-I(0)/I(\Delta_{\phi})$, where $I(0)$ is the intensity at zero-energy and $I(\Delta_{\phi})$ the intensity at the peak (the gap energy). 
Note that for a large gap at $\phi$ there is a large loss of spectral weight,
while $L(\phi)$ vanishes if there are gapless excitations at $\phi$.
In Fig.~\ref{L_vs_Phi} we plot $L(\phi)$ for the same two samples at various $T$'s. Again, we see that there are only small changes in $L$ in the SC state with a fully developed gap right up to $T_c$. 

In the pseudogap regime on the other hand, the data in Fig.~\ref{raw_data} panels {\textbf d} (90K) and {\textbf e} (115K), show that the following dramatic changes occur upon
heating through $T_c$. (1) There are no sharp peaks, implying that long-lived quasiparticle excitations
do \emph{not} exist above $T_c$. (2) While the magnitude of the maximum gap on the zone boundary depends very weakly on $T$, the overall anisotropy of the pseudogap above $T_c$ shows a marked $T$-variation. (3) In particular, the point node in the SC gap below $T_c$ suddenly expands into a Fermi arc of gapless excitations in the pseudogap state above $T_c$.
The transition from point nodes to arcs is clearly visible in the raw data, with a single gapless spectrum at the node at 70 K (Fig.~\ref{raw_data}{\textbf c}) expanding into seven angles with gapless spectra at 90 K (\ref{raw_data}{\textbf d}), and nine angles at 115K (\ref{raw_data}{\textbf e}). Qualitatively similar results are found for the $T_c = 67$ K sample.

To quantify these remarkable changes, we plot the gap in Fig.~\ref{Delta_vs_Phi} and the loss of spectral weight in Fig.~\ref{L_vs_Phi}. 
We see from Fig.~\ref{Delta_vs_Phi} that the energy gap begins to acquire a significant $T$ dependence above $T_c$ in the near nodal region. The point node below $T_c$ expands into a Fermi arc just above $T_c$, which then increases with increasing temperature, as reported earlier \cite{Nature_Physics}. The transition from point nodes to arcs is also clearly visible in the $T$ dependence of the loss of spectral weight $L(\phi)$  in Fig.~\ref{L_vs_Phi}.

\begin{figure}
\begin{center} 
\includegraphics[width=7cm]{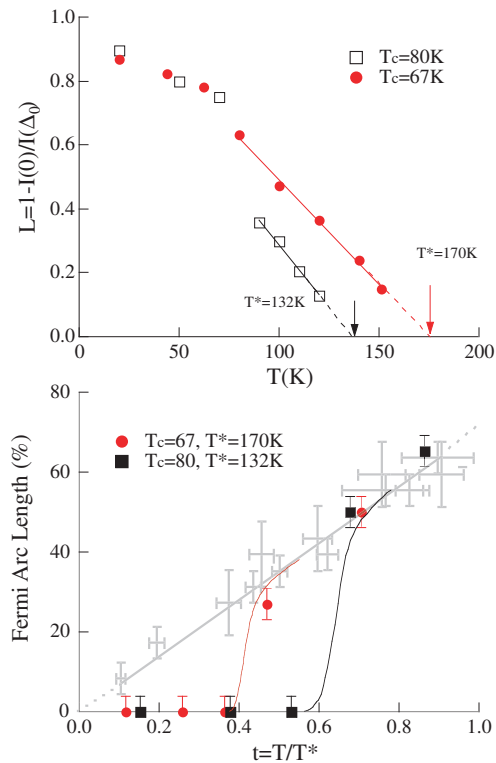}
\end{center}
\caption{Upper panel: loss of zero energy spectral weight at the antinode as a function of temperature for the two samples. The lines are linear fits to the data in the pseudogap phase, the extrapolation to zero yields the value of T$^{*}$ for each sample. Lower panel: Fermi arc length as a function of t=T/T$^{*}$ for both samples. The straight line represents the universal scaling of the arc length found in Ref.~\onlinecite{Nature_Physics}.  The solid curves are the (rescaled) resistivity of the
two samples. }
\label{fig4}
\end{figure}

Close to the antinode the behavior is quite different from the arc region, and the pseudogap does not ``close'' with increasing temperature, showing very little $T$ dependence even above $T_c$.
The evolution of the pseudogap here can be best described as spectral weight 
``filling in'' with increasing $T$, as evidenced by the strong $T$ dependence of the loss function at small $\phi$ in Fig.~\ref{L_vs_Phi}. It is useful to plot the loss of spectral weight at the antinode $L(\phi = 0)$ as a function of $T$; see Fig.~\ref{fig4} (upper panel). In the SC state we see a small, but systematic, decrease
in $L(0)$ with increasing $T$, to which both the drop in $I(\Delta)$ at the peak position (evident in the data
of Fig.~\ref{raw_data}) and the filling in of $I(0)$ contribute. But at $T_c$ there is a sharp, almost 
discontinuous, change of behavior. Above $T_c$, $L(0)$ is entirely dominated by the change in $I(0)$ and the pseudogap fills in linearly with temperature. A simple linear extrapolation of $L(0)$ to zero allows us to determine the temperature $T^{*}$ above which the pseudogap disappears (i.e., has filled in) and a full Fermi surface is recovered.

Finally, we address the sharp collapse of the Fermi arcs at $T_c$. In the lower panel of Fig.~\ref{fig4} we show the length of the Fermi arc as a function of the scaled temperature $t=T/T^{*}(x)$ for the two samples discussed in this paper. We also include here all of the data from Ref.~\onlinecite{Nature_Physics}, where we described the linear scaling of the Fermi arc length with $t$ in the pseudogap state. Our goal here is to see how one \emph{deviates} from the scaling line at $T_c$. We see that, as a function of decreasing $T$ at a fixed doping $x$, there is an abrupt collapse from an arc length which lies on the scaling line above $T_c$ to a point node below $T_c$. The size of the jump depends only on $T_c$: the higher $T_c$ the larger the jump, with the arc length just above $T_c$ simply determined by $T_c/T^*$. We also plot the (rescaled) resistance measured on the same samples \cite{Raffy_T*}, which clearly shows that the abruptness of the collapse of the Fermi arcs has the same width as the resistively measured SC transition.  This implies that the Fermi arc is {\it not} simply due to lifetime broadening of the node.

We now comment on the implications of our results. The observation 
of a $T$-independent SC gap with no sign of the impending phase transition, even just below $T_c$, is very unusual, implying that the energy gap in the underdoped SC state is \emph{not} directly related to the superconducting order parameter.
As already mentioned above, this is qualitatively different from the     
 mean field description of d-wave superconductivity.

For $T < T_c$ let us ask if the essentially $T$-independent gap is consistent with 
nodal quasiparticles, whose existence is well established \cite{Hardy,Chiao}. 
We expect a $T^3$ suppression of the gap  
from thermally excited quasiparticles, given by $\Delta(k,T) = \Delta(T) (\cos k_x - \cos k_y)/2$ with 
$\Delta(T) / \Delta(0) = 1 - \alpha \left(T / \Delta(0)\right)^3$ for $T \ll \Delta(0)$,
with $\alpha$ of order unity.
For the $T_c$ = 80 K sample, $\Delta(0) = 35$ meV and therefore the fractional change in the gap (at any k point) going from 0 K to 70 K is only a few percent. The effect is even smaller for the T$_c$ = 67 K sample, because $\Delta(0) = 42$ meV is larger. Thus thermal effects of nodal quasiparticles are not expected to influence the SC gap, since the scale of the gap $\Delta(0)$ is set by $T^*$ \cite{Campuzano_99}, which is much larger than $T_c$ for underdoped samples. 
   
Some authors argue that the pseudogap originates from a competing order parameter \cite{Ddw,Varma}.
This then leads to a ``two gap'' scenario \cite{Gallais} for the SC state, in which the gap near the antinode is associated with the non-SC order parameter, and scales with the pseudogap  $T^{*}$, while 
the gap which opens up on the arcs is associated with superconductivity, and scales with $T_c$. 
Our results disagree with this scenario in two major ways. First, if this were the case, there would be
no reason to expect a single function of the form $\vert \cos k_x - \cos k_y \vert$  to describe the gap that 
we observe for all $T < T_c$. Second, one would have expected that the small gap near the nodes should have knowledge of $T_c$, and exhibit significant $T$ variation below $T_c$, which it does not; the node (and the gap anisotropy) appears to be protected up to $T_c$.

An alternative explanation, consistent with our underdoped results, is that $T_c$ 
is controlled by the small superfluid density \cite {Uemura,Kivelson_Nature_95} of a doped Mott insulator.
This was first suggested by the Uemura scaling \cite{Uemura} between $T_c$ and the $T=0$ superfluid density 
in underdoped samples \cite{Comment}.
Thus it is fluctuations of the phase of the SC order parameter which destroy superconductivity,
and not the (conventional) collapse of the SC gap. 
The pseudogap is then a remnant of the SC gap above $T_c$ in a state which has no long-range phase coherence \cite{Randeria}. 
This is also consistent with the observation of an unusual Nernst signal \cite{Ong} in a range of temperatures above $T_c$, 
but a detailed theoretical understanding of the pseudogap all the way up to $T^*$ is lacking at the present time. 
 
In summary, we presented a complete description of the thermal evolution of the spectral gap in 
underdoped Bi2212.
The remarkable $T$-independence of the energy gap below T$_c$ implies a non-mean field origin to the SC gap.  And the sudden expansion of
the d-wave nodes present below T$_c$, into finite length Fermi arcs above $T_c$ which scale as 
$T_c  /T^*$, suggests that the Fermi arcs in the pseudogap regime are not simply thermally broadened nodes.

This work was supported by NSF DMR-0606255 and
the U.S. DOE, Office of Science, under Contract No.~DE-AC02-06CH11357.
The Synchrotron Radiation Center is
supported by NSF DMR-0084402.

\end{document}